\documentclass[pre,12pt,preprint,a4paper,showkeys,showpacs]{revtex4}
\usepackage[dvips]{graphics}

\begin{document}

\title{Ergodicity and Central Limit Theorem in Systems with Long-Range Interactions}
\author{A.\ Figueiredo, T.\ M.\ Rocha Filho and M.\ A.\ Amato}
\affiliation{Instituto de F\'\i{}sica, Universidade de Bras\'\i{}lia,
CP: 04455, 70919-970 - Bras\'\i{}lia, Brazil}

\begin{abstract}
In this letter we discuss the validity of the ergodicity hypothesis in theories
of violent relaxation in long-range interacting systems. We base our reasoning on
the Hamiltonian Mean Field model and show that the life-time of quasi-stationary
states resulting from the violent relaxation does not allow the system to reach
a complete mixed state. We also
discuss the applicability of a generalization of the central
limit theorem.
In this context, we show that no attractor exists in distribution space for the
sum of velocities of a particle other than the Gaussian distribution.
The long-range
nature of the interaction leads in fact to a new instance of sluggish convergence to a
Gaussian distribution.
\end{abstract}



\maketitle

\section{Introduction}

Physical systems with long-range interaction has been widely
studied with results reported in the literature, and
properties that are markedly different from short
range interacting systems, e.g, non-Gaussian quasi-stationary states, negative
heat-capacity, temperature jumps at critical points, anomalous diffusion and
ensemble inequivalence~\cite{newproc,oldproc}. Examples include self-gravitating systems,
vortices in two-dimensional fluids, dipolar
interactions, fractures in solids and simplified models as the
Hamiltonian Mean Field (HMF), free electron laser and the plasma single wave
models~\cite{oldproc,chavanis2,sire,hmforig,felmod,tennyson}.
The nature of the non-Gaussian
quasi-stationary states arising from the dynamics can be interpreted as
stable stationary states of the mean-field Vlasov equation which describes
the system in the limit $N\rightarrow \infty$, where $N$
is the number of particles~\cite{r11,r12}.
This equivalence was explicitly verified by
comparing results from direct simulations of a hamiltonian system and the
numerical solutions of the associated Vlasov equation~\cite{antoniazzi}. For
longer times correlations build between particles and the Vlasov equation must be replaced
by a proper kinetic equation~\cite{chavanis,binney}.
The initial stage of time evolution is called violent
relaxation and corresponds to a very rapid evolution from the initial
condition into a quasi-stationary state. Since the seminal paper by
Lynden-Bell~\cite{lyndenbell}, many attempts were made to
formulate the statistical mechanics of violent relaxation (see for instance
References~\cite{arad} and~\cite{18c} and references therein), but a satisfactory theory
for predicting the outcome of the violent relaxation process is still lacking.

In the last decade many papers tried to identify the outcome states of a violent relaxation
with the q-Gaussian distributions:
\begin{equation}
G_q=C\left[1-(1-q)\beta x^2/2\right]^{1/(1-q)},
\label{qgauss}
\end{equation}
arising from the maximization of the Tsallis entropy~\cite{latora,latora2,pluchino,pluchino2}.
Nevertheless some authors pointed out that this approach is too limited in scope, since
q-Gaussian are too specific to handle the richness of quasi-stationary states observed in long-range
interacting systems~\cite{nos1,18b,18c,18d}.
More recently some effort was made to generalize the Central Limit Theorem
(CLT). It was shown that for a special type of correlation (called
q-independence) the sum of stochastic variables tend to a q-Gaussian
distribution \cite{umarov,tsallis1,queiros}. Some attempts were
made to provide specific examples where such type of behavior occur~\cite%
{tirnakli,moyano,pluchino3,pluchino4,25b,25c,ernesto,32}.
In a recent paper Hilhorst and Schehr \cite{hilhorst}
have shown analytically that the sum of correlated stochastic variables for two
such examples (see Ref.~\cite{moyano} and Reference 12 of~\cite{hilhorst})
are in fact not q-Gaussians, but
functions that are only approximated by q-Gaussians. Dauxois also raised
some additional points on the applicability of this approach\cite{dauxois,tsallis2}.
Rodriguez and collaborators described a simple model with scale invariance where this generalized CLT
indeed applies~\cite{rodrigues}.
This shows that examples satisfying its hipothesis
are difficult to obtain. In this work we are interested to study the sum of stochastic variables
given by the momenta of particles in a classical long-range interacting system, the
HMF model~\cite{pluchino3,pluchino4,ernesto}, and possible relationships with the generalization of the CLT, and
to ergodicity.

The HMF model is a paradigmatic and well-studied model, displaying many
characteristics of long-range interacting systems.
In references it is argued, based on numerical simulations,
that the q-generalized CLT
is realized in the sum of velocities of particles of the HMF model.
On this basis it was also argued that the system is non-ergodic.
Pluchino and co-workers refined this assertion by pointing out that
q-Gaussians are only obtained in some realizations of their numerical
simulations (depending on the microscopic details of the initial
conditions), while in the remaining cases the distributions are either
Gaussian or a sort of intermediate function between Gaussians and
q-Gaussians~\cite{32}. In this paper we show that if
an attractor ever exists in the distribution space it is a
Gaussian. The distributions obtained in the conditions used in
references~\cite{pluchino3,pluchino4,ernesto,32} are neither q-Gaussians nor limit
distributions in the sense of a genuine CLT.

We also discuss that the time averages over a single particle history and ensemble averages
are different, albeit diminishing with time. This leads to the conclusion that even-though
the system becomes more ergodic (mixed), the quasi-stationary state relaxes to
the final thermodynamic equilibrium before the ergodicity condition is reached.
This explains the failure of Lynden-Bell theory of violent relaxation, which requires
a complete mixing state.

\section{Central limit theorem and long-range interactions - The HMF model}

Let us start by writing down the hamiltonian for the HMF model~\cite{hmforig}:
\begin{equation}
H=\sum_{i=1}^{N}\frac{p_{i}^{2}}{2}+\frac{1}{2N}\sum_{i,j=1}^{N}\left[
1-\cos (\theta _{i}-\theta _{j})\right] .  \label{hmfham}
\end{equation}
This model describes a system of $N$ rotors with angular positions $\theta_{i}$
and angular momenta (velocities) $p_{i}$.
The magnetization $M$ is defined as the average value of the modulus of the vector
$(\sin\theta,\cos\theta)$:
\begin{eqnarray}
 {\cal M}_x & = & \frac{1}{N}\sum_{i=1}^N\sin\theta_i,\nonumber\\
 {\cal M}_y & = & \frac{1}{N}\sum_{i=1}^N\cos\theta_i,\nonumber\\
 {\cal M} & = & \sqrt{{\cal M}_x^2+{\cal M}_y^2}.
\label{defmag}
\end{eqnarray}

The classical central limit theorem states that the sum of $n$ non-correlated
stochastic variables with finite standard deviation converges to a Gaussian distribution.
The standard deviation of the summed variable scales with $\sqrt{n}$~\cite{levi}. Let us then consider
the discrete time average momentum of a fixed particle in a
system with $N$ particles, say the $k$-th particle:
\begin{equation}
\overline{p}_k=\frac{1}{n}\sum_{i=1}^n p_k(i\Delta t),
\label{vmedia}
\end{equation}
where $\Delta t$ is a fixed time interval. This average is also a stochastic variable
when considering its values for each of the $N$ particles in the system. If $\Delta t$ is sufficiently
large so that the momenta of the same particle at different times are not correlated,
and for statistically stationary states,
the standard deviation scales as $n^{-1/2}$.
Due to the mean-field nature of the system, all relevant physical quantities can be obtained from
the one-particle distribution function $f({\bf r},{\bf p},t)$. If the standard deviation of
$\overline{p}_k$ approaches zero for large enough $n$, then computing averages using
$f({\bf r},{\bf p},t)$ is equal to time-averages over one single particle evolution,
i.~e.\ the system is said to be ergodic.
For the purposes of the discussion below and for comparison with other works, we also define
the reduced average momentum (vanishing average and unit standard deviation) by:
\begin{equation}
y_k=\frac{1}{\sigma}\left[\overline{p}_k-\mu\right],
\label{ykdef}
\end{equation}
where $\sigma$ is the standard deviation of $\overline{p}_k$:
\begin{equation}
\sigma=\sqrt{\frac{1}{N}\sum_{k=1}^N \overline{p}_k^2},
\label{standdevpk}
\end{equation}
and
\begin{equation}
\mu=\frac{1}{N}\sum_{k=1}^N \overline{p}_k
\label{avgdef}
\end{equation}
which vanishes for systems with zero total momentum.
A (central) limit exists
for the sum of stochastic variables if the ensuing distribution has a limit,
or rephrasing, if any finite statistical moment has a limit for $%
n\rightarrow \infty $ which then corresponds to the moment of the limit
distribution. This is the meaning of the word \textit{limit} in the name of
such theorems.
Reduced variables are useful for comparing distribution functions with different averages or
dispersions, but which are of the same type, e.~g.\ all Gaussian distributions are the same
with respect to the reduced variable. Let us note that unfortunately contrary to what is stated in
Ref.~\cite{pluchino3}, ergodicity is not equivalent to requiring that the distribution of each single
$y_k$ and the reduced distribution of the momenta are the same.
As a counter example, consider a system which randomly jumps between two states represented by $-1$ or $+1$.
The state of the system after the $i$-th jump is labeled by the random variable $z_{i}$.
The distribution of probabilities after a large number $n$ of jumps
converge to $P(-1)=P(+1)=1/2$. Since the variables $z_{i}$ are
uncorrelated and have finite standard deviations, the sum
$Z_{i}=\sum_{i=1}^{M}z_{i}$, approaches a Gaussian distribution as $n$
increases, which is obviously different from the distribution for $z_{i}$.
Clearly this does not imply that the system is non-ergodic, as this is
a very simple case of an ergodic system (time averages are equal to
ensemble averages). Therefore arguing that a fixed time reduced distribution
of velocities different from the distribution of the summed variable $y_{i}$
as defined in eq.~(\ref{ykdef}) implies non-ergodicity is not correct.
We show below that for long-range interacting systems the dispersion
in $\overline{p}_k$ scales much slower than $n^{-1/2}$,
even at thermodynamical equilibrium, due to the presence of strong velocity auto-correlations.
If the life-time of the quasi-stationary state is shorter than the time required for $\sigma$
in eq.~(\ref{standdevpk}) to significantly approach zero, then the assumption of ergodicity
used in Lynden-Bell theory breaks down. In the present paper we study this issue in numerical
simulations of the HMF model.

\begin{figure}[tbp]
\hspace{-3mm}\scalebox{0.3}{{\includegraphics{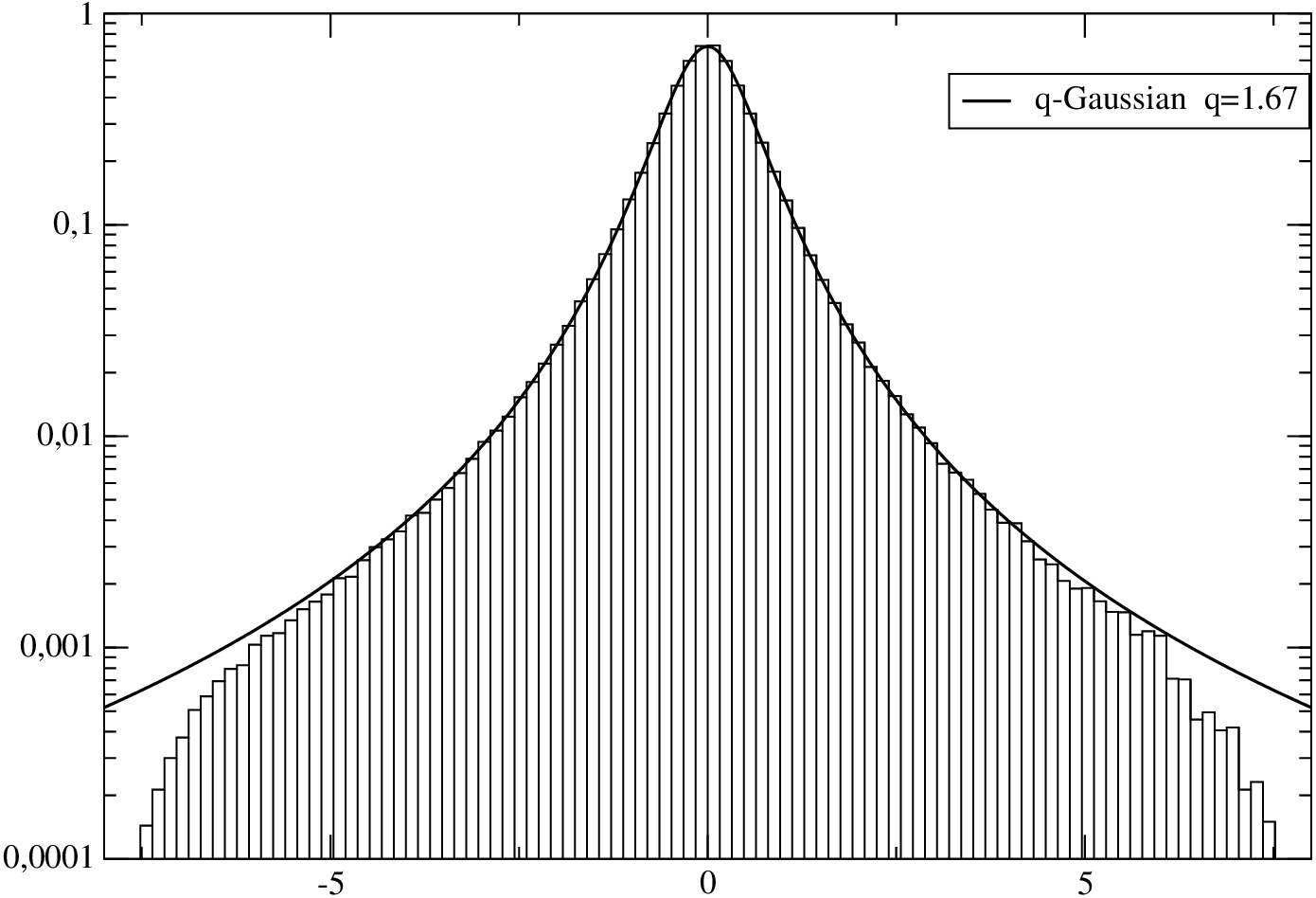}}}\hspace{3mm} %
\scalebox{0.3}{{\includegraphics{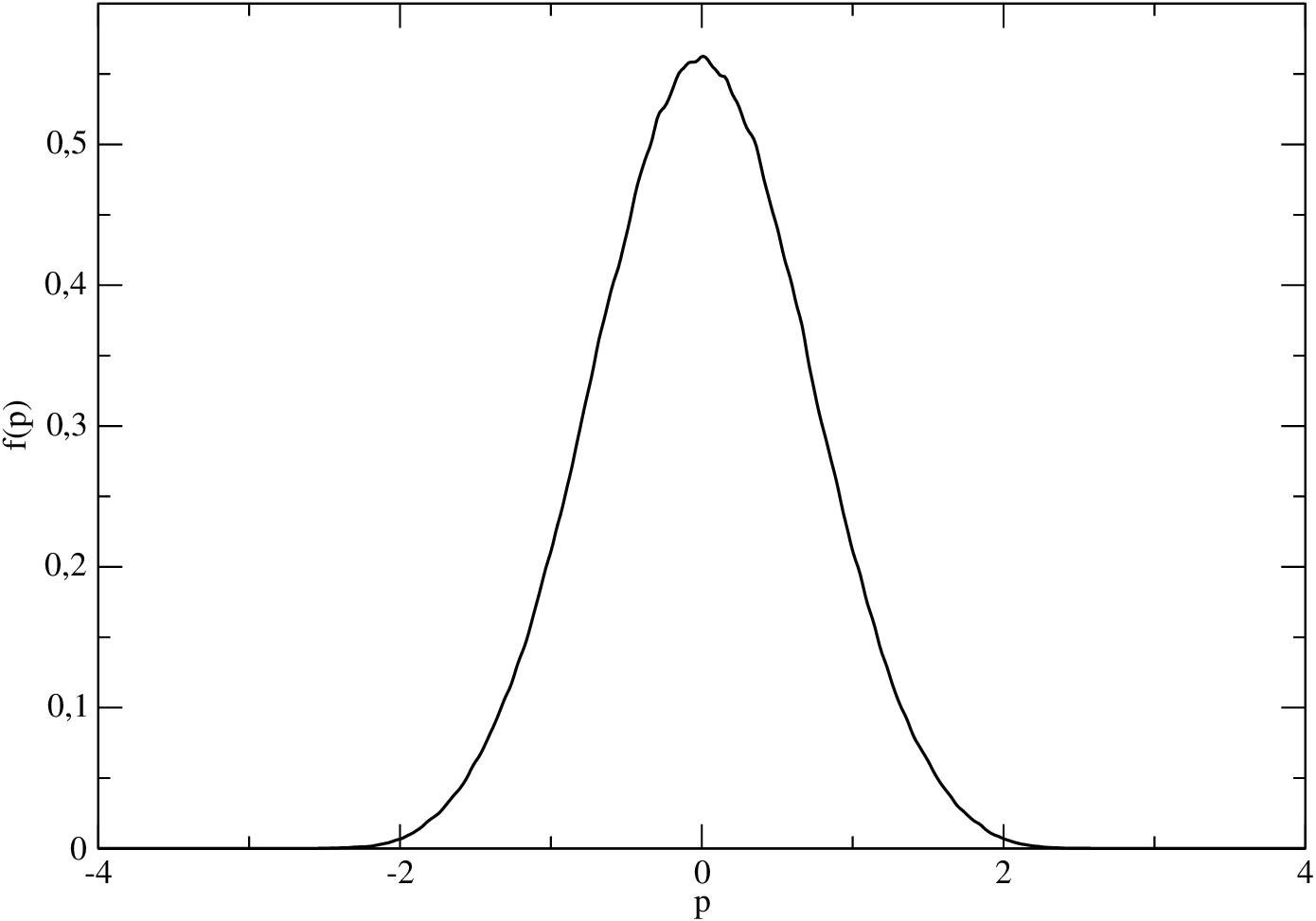}}}
\caption{Left panel: histograms for the distribution of the variable $y_{i}$
over 10,000 realizations, $N=100$, $\Delta t=40$, $E=0.69$ and the best
fitted q-Gaussian with $q=1.67$ and $\protect\beta =5.88$. Right panel:
momentum distribution for $t=2100$.}
\label{fig1}
\end{figure}

We first consider the issue of whether a limit distribution exists for the
reduced variable $y_k$, and if so if it is a q-Gaussian, as claimed in
Refs.~\cite{pluchino3,pluchino4,ernesto,32}.
In order to extend the analysis of distributions we perform numeric
simulations using the parameter values of~\cite{pluchino3,pluchino4,
ernesto} but improving statistical precision. Let us consider
the case studied in Ref.~\cite{pluchino3} with the following parameters:
$N=100$, $\Delta t=40$ and $n=50$, with a transient
time $t_{r}=100$ to allow the system to reach the quasi-stationary state,
with all particles at the origin $\theta _{i}=0$ (magnetization ${\cal M}=1$) and the momenta uniformly
distributed in an interval $-p_{0}<p<p_{0}$,
with an energy per particle $E=0.69$. The
simulations were performed using a fourth-order simpletic integrator~\cite{yoshida},
the same as in~\cite{pluchino3}, and repeated using a different
simpletic integrator for checking~\cite{30b}. The maximum relative energy error
admitted was $10^{-6}$, and we average over 10,000 realizations (10 times
more than in~\cite{pluchino3}), which gives a much better statistical
accuracy in the tails of the distribution. At this point we note that
the use of a good quality random number generator is strongly required to
avoid unwanted correlations when generating the initial conditions. For a
good discussion of this point see~\cite{33,34}. The resulting distribution
for $y_{k}$ is shown in the upper panel of figure~\ref{fig1}, together with
the best fitted q-Gaussian with $q=1.67$ (the value obtained in~\cite%
{pluchino3} is $q=1.65$). It is clear from the tails of the distribution
that it is not a q-Gaussian. The lower panel in figure~\ref{fig1} shows the
reduced velocity distribution function at time $t=2100$ (the final time of the interval over
which the sum~(\ref{ykdef}) runs), which coincides with the one obtained in~\cite{pluchino3}.
This distribution is not Gaussian and can be well approximated as a limit case of a Lynden-Bell distribution
for a two-level initial distribution for ${\cal M}\rightarrow 1$ (see eq.~(\ref{lbpred}) below).
It is straightforward to show that the distribution of $y_k$ cannot be a q-Gaussian
from the fact that for $q>1$ the statistical moments of the reduced variables $y_i$:
\begin{equation}
M_{j}=\frac{1}{N}\sum_{k=1}^{N}\left| y_{k}\right|^{j},
\label{moments}
\end{equation}
diverge for $j>j^{\prime }$ for some $j^{\prime }$. In fact eq.~(\ref{moments}) is an approximation from the simulations
for the true moments obtained for $N\rightarrow\infty$. If the true value of $M_j$ diverges, then computing the simulation value
of $M_j$ from eq.~(\ref{moments}) for increasing $N$ would show no convergence to a finite value.
Nevertheless in all our simulations the value of $M_j$ always converged to
a finite value for a given $j$ such that the corresponding fitted q-Gaussian implied an infinite value.

\begin{figure}
\hspace{0mm}\scalebox{0.3}{{\includegraphics{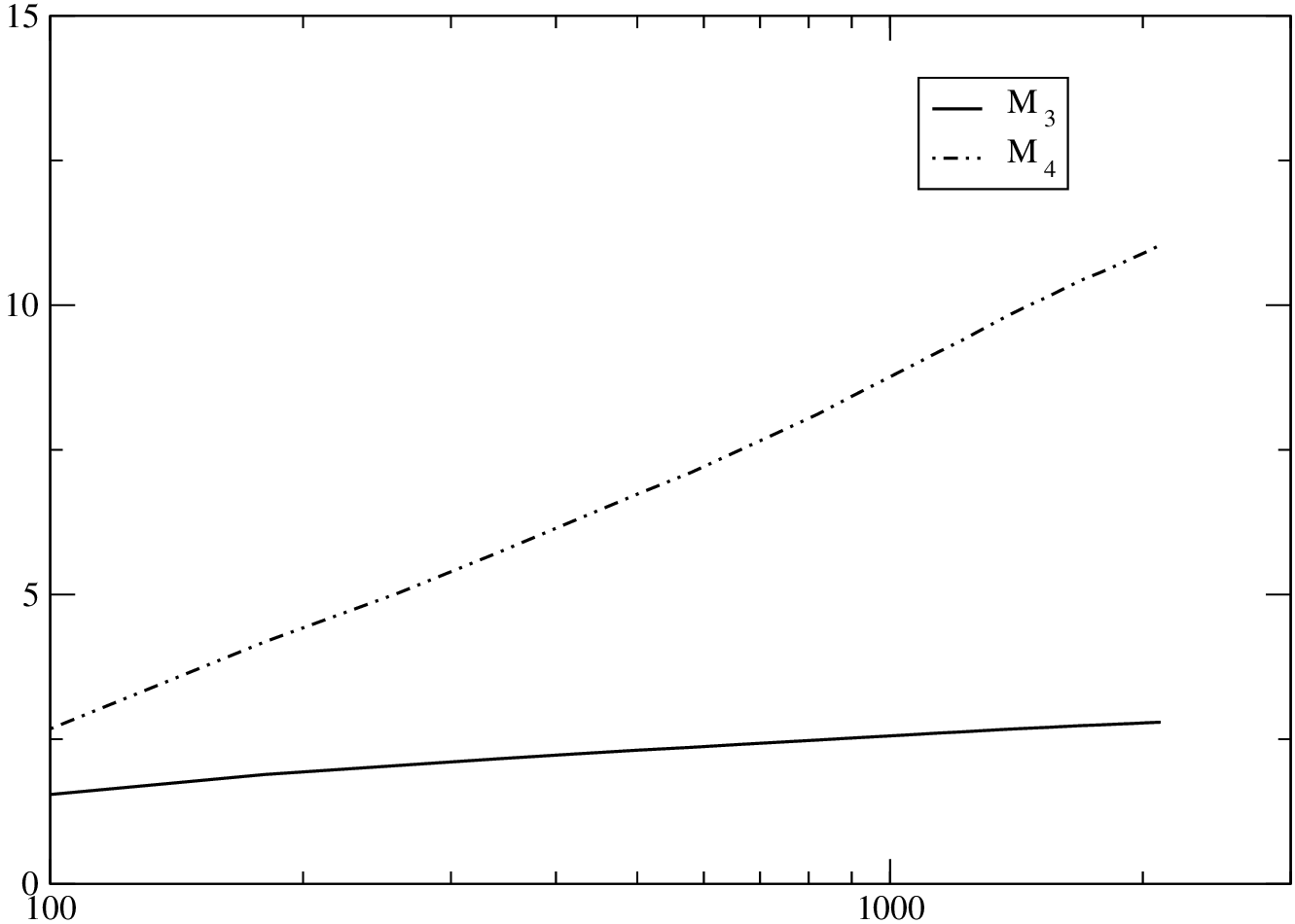}}}
\caption{Moments $M_3$ and $M_4$ of the distribution of the variable $y_k$, for the case in fig.~\ref{fig1}, as a function of $t=n\Delta t$
(varying $n$). Here it is clear that no limit reduced distribution is attained as must be the case for a limit theorem (see comment
after eq.~(\ref{avgdef})).}
\label{fig2}
\end{figure}

Another important point is that the situation at the final time is not a limit state, i.~e.\ the moments
of the reduced variable $y_k$ do not converge to a constant value for increasing time, as seen in fig.~\ref{fig2}.
Along these lines, for the situation considered in~\cite{pluchino3}
no limit distribution is obtained.

\begin{figure}[tbp]
\hspace{0mm}\scalebox{0.3}{{\includegraphics{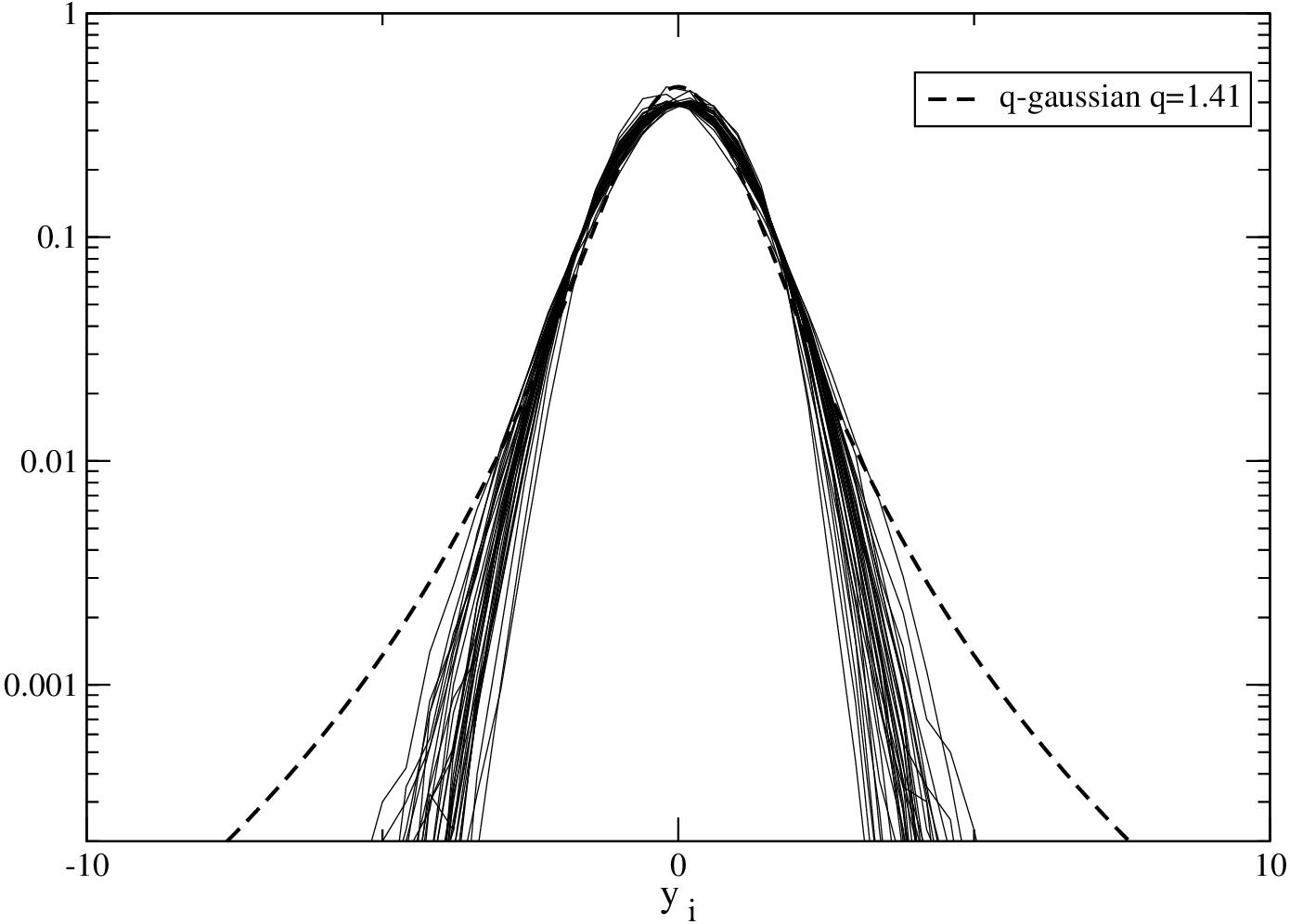}}}\hspace{15mm}
\hspace{0mm}\scalebox{0.3}{{\includegraphics{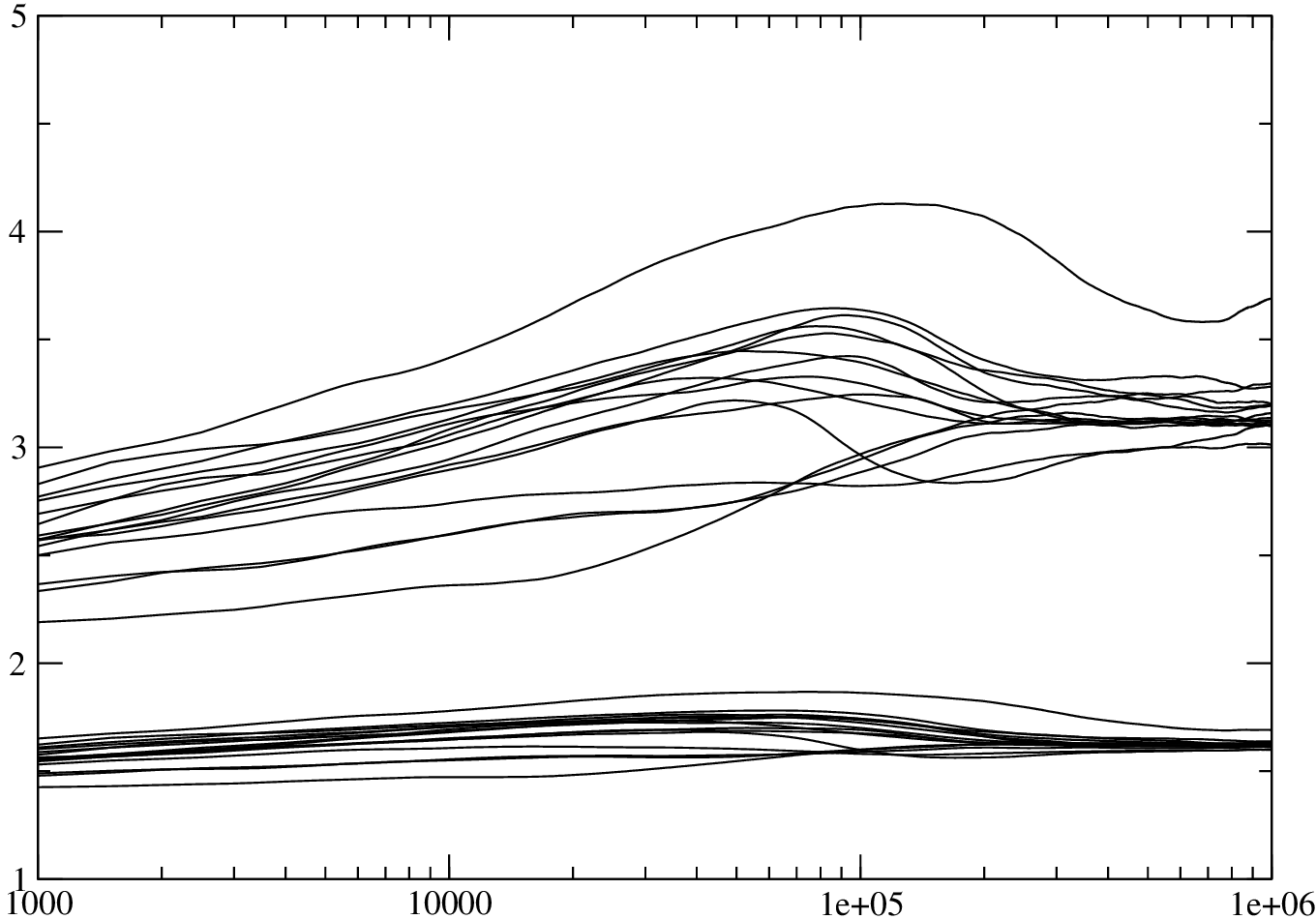}}}
\caption{Left panel: distributions functions for $y_i$ in eq.~(\ref{ykdef}) with $N=100,000$,
$\Delta t=100$ $n=10,000$ and $E=0.69$. Some distributions are very close to a Gaussian.
The best fit using a q-Gaussian (distribution with minimal $\chi^2$
error) is also shown. Right panel: moments $M_3$ (lower set of lines) and $M_4$ (upper set)
as a function of time $t=n\Delta t$
for 15 different realizations showing a sluggish convergence to the values for a Gaussian.}
\label{fig4}
\end{figure}

Extending our analysis let us consider the case studied in ref.~\cite{pluchino4},
with $N=100,000$, $\Delta t=100$, $n=10,000$ and $E=0.69$, with 50 realizations and the same type
of initial conditions. The results
are shown if fig.~\ref{fig4}, together with the q-Gaussian fit with smallest
$\chi ^{2}$ error. It is evident from the figure that no q-Gaussian is realized. The
lower panel in fig.~\ref{fig4} shows the moments $M_{3}$ and $M_{4}$ for 15 different realizations as
function of time.
The reference values of the moments for a reduce variable
having a Gaussian distribution are $M_{3}=1.5958$ and $M_{4}=3$.
The figure shows that all cases, except one which is far from a limit situation,
are converging to a Gaussian distribution.
The present case
is an example of sluggish convergence of a stochastic
process to a Gaussian, as discussed in a different context in References ~\cite{annibal,manteiga}.
The convergence rate itself depends on the details of the initial conditions.

From the results presented above we can draw the following conclusions: the
only true attractor observed in distribution space for the sum in eq.~(\ref{ykdef})
 is the Gaussian distribution. In all cases considered here and in~
\cite{tsallis1,queiros} the distribution is never a q-Gaussian. The only
situation where a limit exists for the distribution of the sum variable is
such that it converges to a Gaussian, as show by the time evolution of
moments. The claim
that a generalized central limit theorem leading to a Tsallis distribution is
observed for the HMF model is therefore invalidated. This do not precludes the
applicability of such theorem in other situations.

\section{Non-ergodic behavior}

\begin{figure}[tbp]
\hspace{0mm}\scalebox{0.4}{{\includegraphics{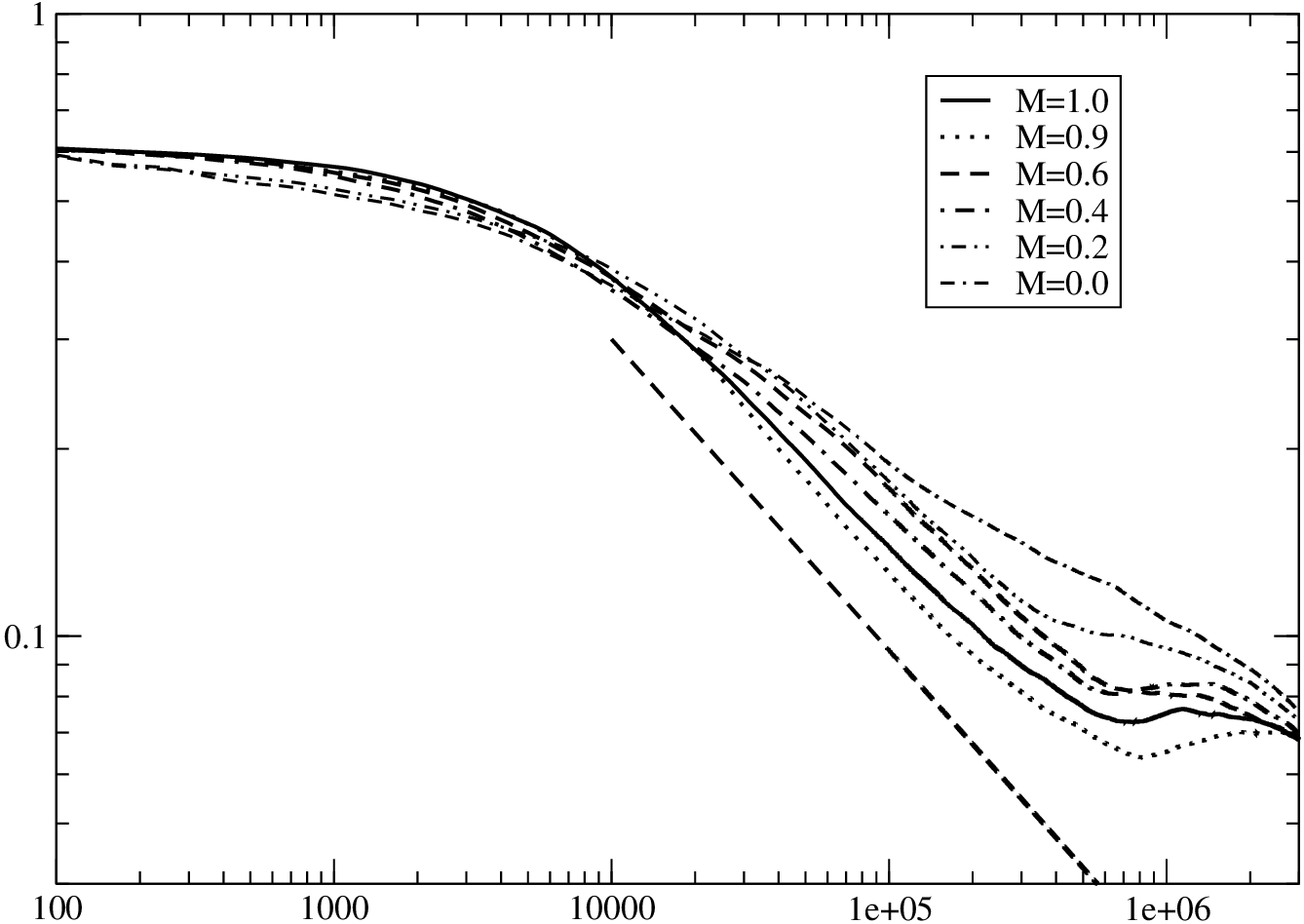}}}\vspace{5mm}
\caption{Dispersion $\sigma$ of the variable $\overline{p}_k$ in eq.~(\ref{standdevpk}) for different
initial magnetizations. The system first evolves 1,000 units of time, allowing the
the violent relaxation to occur and settle down close to a stationary state. The
time indicated in the figure is measured from the end of this transient. The dashed line
is for reference and corresponds to a function proportional to $t^{-1/2}$.
}
\label{fig5}
\end{figure}

\begin{figure}[tbp]
\hspace{0mm}\scalebox{0.4}{{\includegraphics{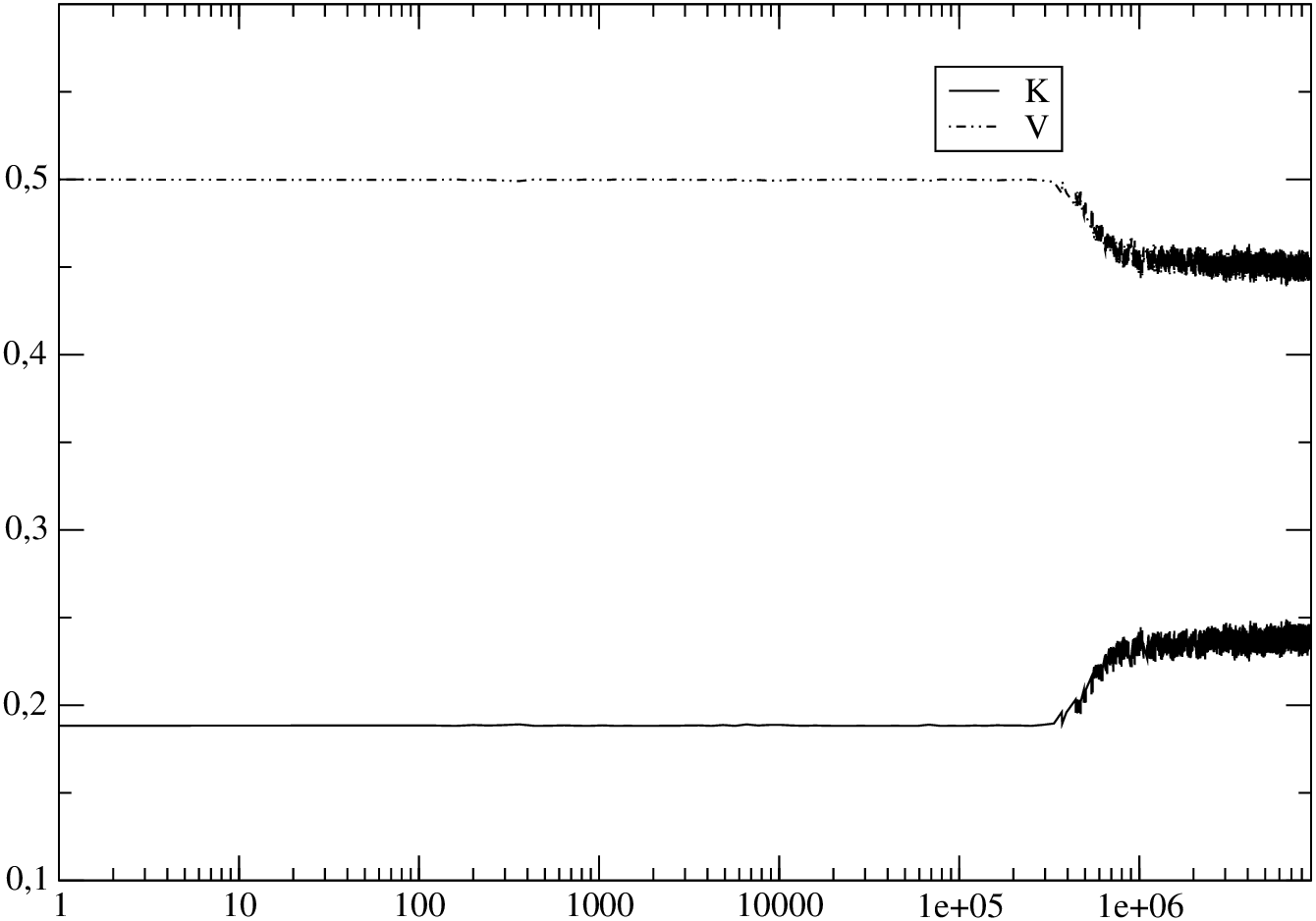}}}\vspace{5mm}
\caption{Kinetic ($K$) and potential ($V$) energies per particle for the case ${\cal M}=0.6$. The transition
from the quasi-stationary state into the final equilibrium (Gaussian) state is given by $\tau=2\times10^5$.
}
\label{fig6}
\end{figure}

\begin{figure}[tbp]
\hspace{0mm}\scalebox{0.4}{{\includegraphics{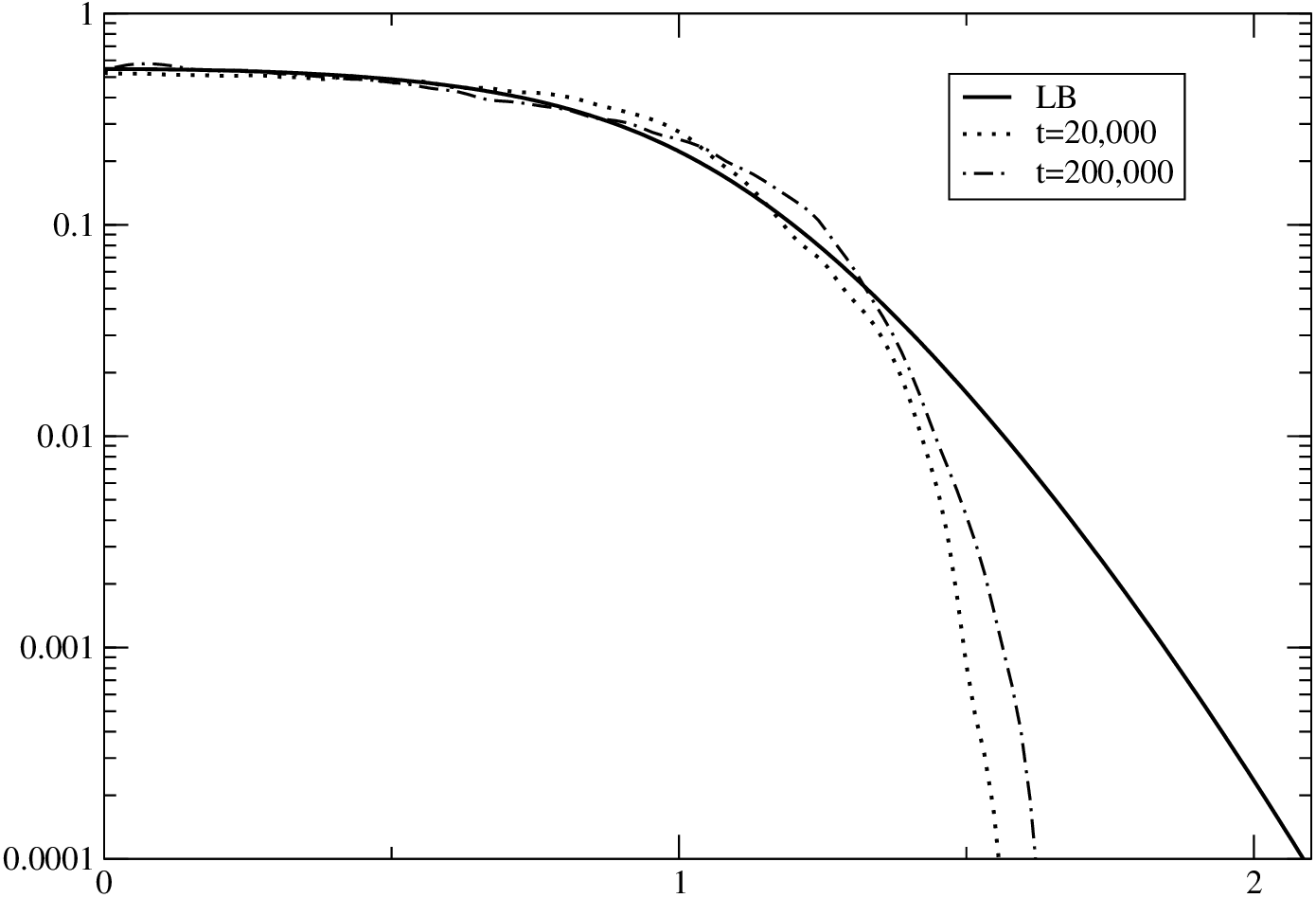}}}\vspace{5mm}
\caption{Distribution functions from simulation for the case ${\cal M}=0.6$ at
two different times compared to the prediction of Lynden-Bell theory. The Lagrange multipliers
obtained by solving eqs.~(\ref{energcond}) and~(\ref{normcond}) are $\beta=4.86$ and $\mu=-6.67$.
}
\label{fig7}
\end{figure}

Now we turn to the ergodicity problem for this system. As explained above, we look at the
time evolution of the dispersion $\sigma$ in eq.~(\ref{standdevpk}). We consider the
HMF model with $N=10,000$ particles, $\Delta t=40$, with a uniform initial distribution
in an interval given by
$f(p,\theta,t=0)=f_0=1/2p_0\theta_0$, if $-p_0<p<p_0$; $0<\theta<\theta_0$, and zero otherwise.
The parameters $p_0$ and $\theta_0$ can be used to adjust the energy and initial magnetization.
The predicted final state being computed using the procedure described in Ref.~\cite{afbcdr}.
In that direction we first note that the value ${\cal M}_x$ in eq.~(\ref{defmag}) in the quasi-stationary state
can be set to zero by a global shift in the angles. The distribution function of the quasi-stationary state
predicted by Lynden-Bell theory for an initial distribution with two levels ($f_0$ and zero) is given by~\cite{lyndenbell}:
\begin{equation}
f_{LB}(p,\theta)=\frac{e^{-\beta e(p,\theta)-\mu}}{1+e^{-\beta e(p,\theta)-\mu}},
\label{lbpred}
\end{equation}
where
\begin{equation}
e(p,\theta)=\frac{p^2}{2}-{\cal M}_y\sin\theta,
\label{onepedef}
\end{equation}
is the one particle energy, and $\beta$ and $\mu$ are Lagrange multipliers determined from
the energy and normalization constraints:
\begin{equation}
\int e(p,\theta) dp d\theta=U,
\label{energcond}
\end{equation}
\begin{equation}
\int f_{LB}(p,\theta)dp d\theta=1,
\label{normcond}
\end{equation}
together with the corresponding expression for ${\cal M}_y$:
\begin{equation}
{\cal M}_y=\int f_{LB}(p,\theta)\sin\theta\:dp d\theta.
\label{myint}
\end{equation}
Equations~(\ref{energcond}) and~(\ref{normcond}) form an algebraic system in the unknowns $\beta$ and $\mu$ and can be solved
numerically to explicitly determine $f_{LB}$. Is is thoroughly discussed in the literature that the predictions of Lynden-Bell
theory yield at best approximate results due to incomplete mixing. We will show that this is in fact due to the strong velocity
auto-correlation which prevents the system to reach an ergodic state, in the sense discussed above.
We show in fig.~\ref{fig5} the dispersion (standard deviation) for the summed variable
$\overline{p}_k$ for some values
of the magnetization of the initial state for $N=10,000$. The life-time of the quasi-stationary state can be inferred
from fig.~\ref{fig6} showing the kinetic and potential energies per particle for the case
${\cal M}=0.6$ (the other cases has essentially the same life-time) as $\tau\approx 2\times10^5$.
The $t^{-1/2}$ scaling for $\sigma$ is reached only after a very long time, of the order of the life-time of the
stationary state. The system evolves to the thermodynamical equilibrium well before the dispersion $\sigma$ can attain
a significantly small value. Figure~\ref{fig7} shows the velocity distribution function for two times compared with the
Lynden-Bell distribution function for ${\cal M}=0.6$. Even though the distribution is slowly approaching the theoretical prediction
while $\sigma$ decreases, the difference remains always significant.


\section{Concluding Remarks}

We have shown that the HMF model is non-ergodic due to strong auto-correlations in particle velocities, which are
also responsible for anomalous diffusion and Levi walks, restraining the dispersion in the time average of the velocity
of each particle to approach zero before the system finally settles down in its thermodynamical equilibrium.
The system has not enough time to reach ergodicity and therefore any intermediate state before final equilibrium
is incompletely relaxed. Ergodicity is only fulfilled for asymptotic times, well after thermodynamic equilibrium occurs.
This clearly points out that a solution of the violent relation problem must necessarily rely on a dynamical theory,
based for instance in the solution of a proper kinetic equation as already pointed out by Chavanis~\cite{18c}.
Many of the points presented here can be extended to other systems with long-range interactions as long as strong
velocity auto-correlations are present. We have also shown, if simulations are performed with care to statistical
precision, that the only attractor observed for the sum of the velocity of a single particle at fixed time intervals
is the Gaussian distribution, thus invalidating the claims by Tsallis and collaborators that such attractors should be
q-Gaussian distributions.

\section{Acknowledgments}

The authors thanks C.~Tsallis for useful discussions. This work was partially financed by CNPq (Brazil).


\begin{thebibliography}{99}
\bibitem{newproc} {\it Dynamics and Thermodynamics of Systems With Long Range
Interactions: Theory and Experiments},
Eds.\ A.~Campa, A.~Giansanti, G.~Morigi and F.~F.~Labini,
AIP Conference Proceedings, Vol 970, (Melville, 2007).

\bibitem{oldproc} {\it Dynamics and Thermodynamics of Systems with Long Range Interaction},
Eds.\ T.~Dauxois, S.~Ruffos and M.~Wilkens, {Lecture Notes in
Physics} Vol {602} (Berlin, 2002).

\bibitem{hmforig} {M.~Antoni and S,~Ruffo}, Phys.\ Rev.\ E \textbf{52} (1995)
2361.

\bibitem{tennyson} {J.~L.~Tennyson, J,~D.~Meiss and P.~J.~Morrison}, Physica D 
\textbf{71} (1994) 1.

\bibitem{chavanis2} {P.~H.~Chavanis}, Phys.\ Rev.\ Lett.\ \textbf{84} (2000) 5512.

\bibitem{sire} {P.~H.~Chavanis and C.~Sire}, Phys.\ Fluids.\ \textbf{13} (2001) 71804.

\bibitem{felmod} {R.~Bonifacio, F.~Casagrande, G.~Cerchioni, L.~De Salvo
Souza, P.~Pierini and N.~Piovella}, Riv.\ Nuovo Cimento \textbf{13} (1990) 1.

\bibitem{r11} {W.~Braun and K.~Hepp}, Commun.\ Math.\ Phys.\ \textbf{56} (1977) 101.

\bibitem{r12} {P.~H.~Chavanis}, Physica A {\bf 361} (2006) 55.

\bibitem{antoniazzi} {A.~Antoniazzi, F.~Califano, D.~Fanelli and
S.~Ruffo}, Phys.\ Rev.\ Lett.\ \textbf{98} (2007) 150602.

\bibitem{chavanis} {P.~H.~Chavanis} Physica A {\bf 387} (2008) 1504.

\bibitem{binney} {J.~Binney and S.~Tremaine}, {\it Galactic Dynamics}
Princeton Univ.\ Press, (Princeton, 1987).

\bibitem{lyndenbell} {D.~Lynden-Bell}, MNRAS \textbf{136} (1967) 101.

\bibitem{arad} {I.~Arad and P.~H.~Johanson}, MNRAS \textbf{363} (2005) 252.

\bibitem{18c} {P.~H.~Chavanis}, Physica A \textbf{365} (2006) 102.

\bibitem{latora} {V.~Latora, A.~Rapisarda and C.~Tsallis}, Phys.\ Rev.\ E 
\textbf{64} (2001) 056134.

\bibitem{latora2} {V.~Latora, A.~Rapisarda and C.~Tsallis C.}, Physica A \textbf{305} (2002) 129.

\bibitem{pluchino} {A.~Pluchino, V.~Latora and A.~Rapisarda}, Physica D \textbf{193} (2004) 315.

\bibitem{pluchino2} {A.~Pluchino, V.~Latora and A.~Rapisarda}, Cont.\ Mech.\ Therm.\ \textbf{16A.} (2004) 245.

\bibitem{nos1} {T.~M.~Rocha Filho, A.~Figueiredo and M.~A.~Amato}, Phys.\ Rev.\ Lett.\ \textbf{95} (2005) 190601.

\bibitem{18b} {P.~H.~Chavanis}, Physica A \textbf{359} (2006) 177.

\bibitem{18d} {C.~F\'eron and J.~Hjorth}, astro-ph/0801.2504.

\bibitem{umarov} {S.~Umarov, C.~Tsallis and S.~Steinberg}, Milan J.\ Math.\ {\bf 76} (2008) DOI: 10.1007/s00032-008-0087-y. cond-mat/0603593.

\bibitem{tsallis1} {C.~Tsallis and S.~M.~D.~Queiros},
AIP Conf.\ Proc.\ \textbf{965} (2007) 8.

\bibitem{queiros} {S.~M.~D.~Queiros and C.~Tsallis},
AIP Conf.\ Proc.\ \textbf{965} (2007) 21.

\bibitem{tirnakli} {U.~Tirnakli, C.~Beck  and C.~Tsallis}, Phys.\ Rev.\ E \textbf{75} (2007) 040106(R).

\bibitem{moyano} {L.~G.~Moyano, C.~Tsallis, and M.~Gell-Man}, Europhys.\ Lett.\ 
\textbf{73} (2006) 813.

\bibitem{25b} {C.~Tsallis}, Physica A \textbf{365} (2006) 7.

\bibitem{25c} {J.~A.~Marsh, A.~M.~Fuentes, L.~G.~Moyano and C.~Tsallis},
Physica A \textbf{372} (2006) 183.

\bibitem{pluchino3} {A.~Pluchino, A.~Rapisarda and C.~Tsallis}, Europhys.\
Lett.\ \textbf{80} (2007) 26002.

\bibitem{pluchino4} {Pluchino A. \and Rapisarda A.} cond-mat/0712.2539.

\bibitem{ernesto} {C.~Tsallis, A.~Rapisarda, A.~Pluchino and E.~P.~Borge},
Physica A \textbf{381} (2007) 143.

\bibitem{32} {A.~Pluchino, A.~Rapisarda and C.~Tsallis}, cond-mat/0801.1914.

\bibitem{hilhorst} {H.~J.~Hilhorst and G.~Schehr}, J.\ Stat.\ Mech.\ (2007) P06003.

\bibitem{dauxois} {T.~Dauxois}, J.\ Stat.\ Mech.\ (2007) N08001.

\bibitem{tsallis2} {C.~Tsallis}, (2007) cond-mat/0712.4165v1.

\bibitem{rodrigues} {A.~Rodriguez, V.~Schw\"ammle and C.~Tsallis } cond-math.stat-mech/0804.1488v1.

\bibitem{levi} {D.~Dugu\'e}, {\it Oeuvres de Paul L\'evy: El\'ements Al\'eatoires} {\bf III}
Gauthiers-Villars (Paris, 1976).

\bibitem{yoshida} {H.~Yoshida H.}, Phys.\ Lett.\ A \textbf{150} (1990) 262.

\bibitem{30b} {I.~P.~Omelyan, I.~M.~Mryglod and R.~Folk}, Comp.\ Phys.\ Comm.\ 
\textbf{146} (2002) 188.

\bibitem{33} {W.~H.~Press, S.~A.~Teukolsky, W.~T.~Vetterling and
B.~P.~Flannery}, {\it Numerical Recipes, 2nd Ed}, Cambridge Univ.\ Press (Cambridge, 1992).

\bibitem{34} {D.~E.~Knuth}, {\it The Art of Computer Programming, 3rd Ed},
Vol.\ {2} Addison Wesley (Reading, 1997).

\bibitem{annibal} {A.~Figueiredo, I.~Gleria, R.~Matsushita and S.~Silva},
Physica A \textbf{337} (2004) 369.

\bibitem{manteiga} {R.~N.~Mantegna H.~E.~Stanley}, Phys. Rev. Lett. 
\textbf{73} (1994) 2946.

\bibitem{afbcdr} {A.~Antoniazzi, J.~Barr\'e, P;~H.~Chavanis, T.~Dauxois and S.~Ruffo},
Phys.\ Rev.\ E {\bf 75} (2007) 011112.

\end{thebibliography}
\end{document}